# When does AI support thinking, and when does it replace it? Learners' conceptualisations of AI as a dynamic cognitive partner: A typology

Cecilia Ka Yuk Chan

Affiliation: The University of Hong Kong

Email: ckchan09@hku.hk

Website: https://tlerg.talic.hku.hk/

https://aiedlab.hku.hk/

## Abstract

Artificial intelligence (AI) is increasingly embedded in education, raising a fundamental question: when learners use AI, does it support their thinking or replace it? While existing research has focused on system capabilities and challenges and opportunities, less is known about how learners themselves conceptualise AI's role in their thinking.

This study examines learners' own accounts of AI use to understand how they position AI within their cognitive processes. Using qualitative analysis of written responses from 145 secondary students (aged 14–17) in Hong Kong, a learner-informed typology that conceptualises AI as a dynamic cognitive partner whose role shifts across learning situations was developed.

The analysis identifies nine interrelated cognitive functions through which learners describe engaging with AI, including conceptual scaffolding, feedback, idea generation, organisation, adaptation, monitoring, and workload regulation. Crucially, across these functions, students consistently distinguish between AI use that extends cognition and AI use that replaces cognitive effort. This reveals a central boundary in AI-supported learning: the same interaction can either support sense-making or enable cognitive offloading, depending on how learners position AI in the learning process.

Grounded in Sociocultural Theory, Distributed Cognition, Self-Regulated Learning, and Cognitive Load Theory, the typology reframes AI not as a fixed instructional tool but as a shifting form of cognitive mediation. By foregrounding the boundary between cognitive extension and substitution, the study provides a conceptual lens for understanding when AI supports learning and when it risks undermining it.

## Introduction

Across educational contexts, AI systems are increasingly used for tutoring, feedback generation, adaptive learning, content organisation, and well-being support (Chan, 2025a; Demartini et al., 2024; Meyer et al., 2024; Zong & Yang, 2025). Recent studies highlight both the opportunities and challenges AI introduces into learning environments with concerns regarding overreliance, cognitive offloading, and diminished learner agency (Chan & Hu, 2023; Giannakos et al., 2025). As generative AI (GenAI) can interact with learners during the act of thinking itself, AI is no longer only providing outputs; it is participating in cognition in action (Essel et al., 2024). This creates a new learning dynamic, co-constructive process in which human and artificial contributions are intertwined.

While existing research questions have examined AI's instructional effectiveness, system design, ethical considerations, learners' acceptability and AI literacy development (Chan, 2025b; Chiu, 2026; Ng et al., 2024), a key question remains underexplored:

> ***How do learners themselves cognitively conceptualise AI's role within their own thinking processes?***

Put differently, when students use AI to learn, how do they position the AI tool in the learning process, as a resource they actively steer to support their reasoning, as a tutor that guides them step-by-step, or as an answer generator that completes work with minimal learner input? Understanding this question is critical because learners' mental models of AI shape how the technology is used, regulated, and integrated into learning process (Liu et al., 2025). If AI is perceived as a shortcut or answer-provider, it may lead to cognitive outsourcing. If it is viewed as a support for thinking, it may function as scaffolding that enhances self-regulated learning (SRL) and conceptual development (Gerlich, 2025; Watson & Ennion, 2026; Wu & Chiu, 2025). Thus, AI's educational impact depends not only on system capability and outputs but also on how learners position AI within their cognitive processes.

The present study addresses this by analysing student written responses in which learners describe how AI supports their learning. These texts capture learners' own perspectives and experiences on moments of explanation, feedback, organisation, adaptation, monitoring, and persistence after their AI literacy course. The analysis focuses not on what AI systems are designed to do, but on how learners describe AI as participating in their thinking across different learning situations.

From the findings, the study derives a typology conceptualising AI as a Dynamic Cognitive Partner. The term *dynamic* reflects that AI's role is not static; its function shifts according to learner needs, task demands, and stages of understanding. It also captures AI's interactive, personalised, and adaptive nature. The term *cognitive partner* emphasises that AI operates within thinking processes, supporting reasoning, sensemaking, monitoring, and regulation, rather than solely at the level of content delivery or administrative assistance.

By grounding the typology in learner perspectives and linking it to established theories of mediated cognition and self-regulated learning, the study offers a process-oriented model for understanding AI's role in human learning activity. Crucially, it identifies a central boundary between AI use that extends cognition and use that replaces it, providing a conceptual lens for understanding when AI supports learning and when it risks undermining it.

## Theoretical Foundations for conceptualising AI as a Dynamic Cognitive Partner

Positioning AI as a *Dynamic Cognitive Partner* requires moving beyond descriptions of AI as a "tool" that delivers content or automates tasks, and instead treating AI as part of the learner's cognitive activity system, a partner that learners recruit to think with, not only to get answers. This section synthesises theory and AI-in-education research to explain how a learner-informed typology is grounded in (a) Sociocultural theories of mediated learning, (b) Distributed Cognition and Extended Mind Thesis, and (c) Self-Regulated Learning and Cognitive Load. Together, these perspectives explain both the promise of AI as cognitive support and the risk that AI can displace learner processing if used as a substitute for thinking.

**Sociocultural Perspective: ZPD and tool-mediated learning**

From a sociocultural perspective, learning is not only an internal cognitive process; Vygotsky (1978) argued that human thinking does not develop in isolation but is mediated through interaction and participation in socially organised activity with cultural tools that are created and shared within a culture. These mediational means include both physical tools (e.g., books, calculators, digital technologies) and psychological tools (e.g., language, symbols, and strategies), all of which influence how learners reason and make sense of the world. Central to this perspective is the Zone of Proximal Development (ZPD), defined as the distance between what learners can accomplish independently and what they can achieve with guidance. Traditionally, such guidance is provided through scaffolding by teachers or peers, characterised by contingency, gradual fading, and transfer of responsibility to the learner (van de Pol et al., 2010). However, scaffolding can become counterproductive "crutches" if learners fail to internalise supported processes (Meegan & Young, 2025). In AI-mediated contexts, learners increasingly describe AI as providing forms of scaffolding—offering explanations, step-by-step guidance, and alternative ways of understanding (Cai et al., 2025). This suggests that learners may position AI as a source of contingent support during conceptual difficulty. At the same time, it is important to note that AI lacks the intentionality and social awareness associated with a traditional "more knowledgeable other," and thus its role is better understood as mediating rather than directing learning.

A key conceptual move is shifting the lens from "AI as a tool" to "AI as cognitive mediation." It is reasonable to be cautious with the phrase *mediational tool,* because AI is not only a static artifact (like a textbook); it can behave like an interactive partner through conversation, iterative refinement, and responsive explanation (Guo et al., 2025; Casheekar et al., 2024). This aligns more closely with the idea of mediated action rather than passive tool use.

Activity theory (Engeström, 1987) extends this view by situating learning within goal-directed activity systems in which tools mediate the relationship between a learner (subject) and a learning goal (object), within broader structures of rules, community, and division of labour. In such systems, AI does not simply provide information but influences how tasks are carried out. For example, when a student uses an AI chatbot during essay writing, the student may ask AI to brainstorm ideas, the system supports idea generation; if the student requests step-by-step clarification of a concept, AI supports conceptual scaffolding; if the student uses AI to check grammar, it functions as a feedback partner. The AI's role therefore shifts depending on the moment-to-moment demands of the task (Li et al., 2025).

When students use AI, they do not simply "access information"; they reorganise the learning activity itself. They offload certain sub-tasks (e.g., summarising), open new cognitive pathways (e.g., rapid brainstorming), and alter temporal rhythms (e.g., immediate feedback instead of delayed teacher marking). These changes help explain why learners describe AI roles as dynamic, situational and adaptive: AI's contribution varies depending on task type, time pressure, confidence level, and the availability of human support (Chan & Hu, 2023; Bai & Wang, 2025). In this sense, AI participates in the mediation of thinking within activity, rather than functioning as a neutral external aid.

Sociocultural theory does not assume that tools automatically enhance learning. The benefits depend on how learners appropriate and internalise mediated processes. Research on AI in education reflects this tension, highlighting both opportunities for conceptual support and risks such as overreliance or reduced critical engagement (Zhai et al., 2024; Chan & Colloton, 2024). From this perspective, AI may support development within the ZPD, but whether it supports development depends on how learners engage with it.

**Distributed Cognition and Extended Mind**

Distributed cognition perspectives argue that cognition is not confined to an individual mind; it includes interactions with tools, representations, and environments (Hutchins, 1995). In education, this view has been used to explain how learners think with diagrams, calculators, notes, digital systems, and peers. AI fits naturally into this tradition: students can use AI to transform messy information into structured representations (outlines, flashcards), to generate alternative explanations, or to externalise intermediate reasoning steps (Sousa & Cardoso, 2025). In distributed cognition terms, AI becomes part of the learner's cognitive system, helping coordinate processes such as remembering, organising, checking, and sensemaking.

The extended mind thesis similarly argues that external resources can become functionally integrated into cognition when they are reliably available, easily accessible, and tightly coupled with the user's goals (Clark & Chalmers, 1998). Modern days AI systems are always available, conversational, multi-purpose demonstrates supports the theoretical angle of the "cognitive partner" framing: if learners repeatedly use AI to scaffold explanations, test understanding, and regulate progress, AI is not simply a peripheral tool but a *functional component* of learners' cognitive workflow.

However, this perspective also highlights a key risk: when external resources become too powerful or too convenient (Rivera-Novoa & Duarte Arias, 2025), learners may rely on them in ways that reduce internal processing. This is not inherently negative, offloading routine tasks can free cognitive resources for deeper work but it becomes problematic when learners offload the very processes that *constitute learning* (e.g., generating explanations, monitoring comprehension, and practicing retrieval) (Gerlich, 2025).

**Self-Regulated Learning and Cognitive Offloading: AI as a regulation partner**

Self-regulated learning (SRL) conceptualises learning as cycles of planning, monitoring, strategy use, and reflection (Zimmerman, 2002). AI can support each of these phases, for example, by helping learners plan tasks, monitor performance through feedback, apply strategies through guided practice, and reflect on progress. However, SRL emphasises that effective learning depends not simply on the availability of support, but on learners' ability to use it strategically (Karabenick & Berger, 2013). Learners who know when and who to seek help, how to evaluate it, and when to persist independently are more likely to benefit.

Empirical research suggests that AI-supported learning is shaped by learners' motivation, prior knowledge, and regulatory capacity (Chiu et al., 2023; Xia et al., 2023). This indicates that AI does not determine outcomes directly; rather, its effects are mediated by how learners engage with it.

Cognitive load theory (CLT) provides a complementary explanation for why learners *want* AI support and why over-reliance can happen. CLT argues that working memory is limited and learning is constrained when tasks impose excessive load (Sweller, 1994). AI can reduce load by summarising, chunking tasks, generating worked examples, and translating complex language into simpler forms. At the same time, CLT warns that when supports do too much of the cognitive work, learners may not engage in the processing needed for schema construction. As Risko and Gilbert (2016) argue, cognitive offloading, the use of external actions or tools to ease cognitive load, can enhance task performance but may also risk dependency. Integrative frameworks connecting cognitive load and SRL emphasise that effective learning requires not just reduced load, but learners' active regulation of attention, strategy, and effort (de Bruin et al., 2020). AI can help regulate load and sustain progress, but the learner's role as the *epistemic agent* must remain intact.

A key SRL mechanism is metacognitive monitoring: learners must judge what they know, identify gaps, and select appropriate strategies. AI can assist monitoring through feedback, explanations, and comparison across solutions; yet AI can also degrade monitoring if learners accept outputs uncritically, leading to epistemic and metacognitive laziness (sloth) (Fan et al., 2025; Yurt & Kuşci, 2026; Zhan & Yan, 2025). SRL and CLT together predict this vulnerability: under time pressure and high workload, learners may prioritise efficiency over learning quality, especially if the AI tool can provide fluent answers quickly. Cognitive load theory explains that while AI reduces extraneous load by simplifying tasks, it may also diminish germane load, that is, cognitive resources directed toward processing and automating information in long-term memory, encouraging learners to invest less effort in schema-building and reflective monitoring as AI provides ready-made answers (Rind, 2026). This dual potential aligns with the idea that AI can either extend cognition or outsource it.

**The Present Study**

Taken together, sociocultural theory, distributed cognition, self-regulated learning, and cognitive load theory converge on a common insight: learning with AI is not simply a matter of access to information, but a matter of how AI becomes integrated into the learner's cognitive system. Across these perspectives, AI can function as scaffolding within the learner's Zone of Proximal Development, as a mediating artifact in goal-directed activity, as an external component of distributed cognition, and as a resource for planning, monitoring, and regulating effort. At the same time, these theories jointly predict a central tension: the same features that make AI a powerful support for understanding can also make it a mechanism for cognitive substitution when learners offload sensemaking, monitoring, or strategy use to the system. Thus, AI's educational value is not inherent in the technology itself but emerges from the patterns of interaction between learner, task, and AI within specific learning situations.

What remains underdeveloped in existing work is a learner-centred account of how students themselves perceive and position AI within these cognitive processes. While theoretical and empirical studies describe AI's affordances and effects, fewer studies reconstruct how learners articulate AI's role in moments of difficulty, idea generation, organisation, regulation, and persistence. Understanding these learner conceptualisations is critical because they shape when AI is used as a scaffold, when it becomes a shortcut, and how responsibility for thinking is negotiated. The present study therefore turns to students' written responses to examine how learners describe AI's participation in their thinking. From these accounts, we derive a learner-informed typology that conceptualises AI as a Dynamic Cognitive Partner, not a static tool, but an adaptive form of cognitive mediation whose role shifts across tasks, needs, and stages of understanding. It needs to be emphasised that the scope of this investigation was limited to examining students' reported perceptions of how they engaged with AI in their thinking processes rather than direct observation of their actual behaviours.

## Methodology

### Participants and Context

Participants were 145 secondary school students (aged 14–17) from five schools across Hong Kong who had completed an online AI literacy course prior to data collection. Schools enrolled in the programme voluntarily, and students participated in a non-graded learning environment that was not part of formal school assessment. This provided a low-stakes context in which students could engage openly with questions about AI use in learning.

Participants were recruited through convenience sampling. Invitations were sent to all students enrolled in the course, and participation in the study was voluntary. Students were informed of the study's purpose, the voluntary nature of participation, and the anonymised use of their responses. Ethical approval was obtained prior to the study. Participation followed school guidelines, and student submissions were de-identified before analysis. All data were stored securely, and no identifying information is reported.

All participants had basic digital literacy and introductory experience with AI tools typical of middle secondary learners in Hong Kong. This included familiarity with online learning platforms, foundational computational thinking (e.g., basic coding), and initial experience interacting with AI systems such as chatbots.

The AI literacy course was a 20-hour online programme comprising five modules covering topics such as basic AI concepts, applications in education, prompt engineering, assessment and AI detection, chatbot creation, and ethical considerations. There were many practical exercises within the course allowing students to get hands-on experience with many AI tools. Students accessed the course between November 2024 and June 2025. Data collection was embedded within the course as a structured learning activity following instruction on AI use and ethics, ensuring that participants had both conceptual understanding and practical experience with AI prior to producing their written responses.

### Data Source

The dataset comprised of a non-graded component - student written responses produced as final part of AI literacy course. Students were instructed to write in English (approximately 400–800 words) describing how AI could support their learning. The task prompted students to share their personal perception and experience in the following areas:

- What AI is and how it may be used in education
- Useful AI tools and how they support learning (e.g., tutoring, feedback, organisation)
- How AI may change the roles of teachers and students
- Challenges or risks of using AI (e.g., inaccuracy, overreliance, integrity)
- How and when the student uses AI to support learning

Students were not describing tools but articulating their own understandings of AI's role in thinking, studying, and problem solving. Importantly, many students also discussed situations where AI might lead to overreliance, reduced effort, or diminished critical thinking, providing data on both supportive and problematic forms of AI engagement.

**Data Analysis**

Texts were analysed using qualitative content analysis with both inductive and theory-informed elements. The focus of analysis was not on specific AI platforms, but on the cognitive functions students attributed to AI.

Two researchers independently coded all texts. The first cycle of coding used open coding to mark segments where students described:

- what AI helps them do cognitively (e.g., explain, check, organise, give ideas),
- when they turn to AI (e.g., when stuck, overloaded, unsure), and
- how they positioned their own role relative to AI (e.g., thinking partner vs. shortcut).

Codes were then compared and clustered into function-oriented categories such as conceptual explanation, feedback, organisation, adaptation, monitoring, and persistence support. During this stage, segments describing risks of AI use, including overreliance, cognitive offloading, or loss of independent thinking, were also coded to capture learners' awareness of boundaries between support and substitution. This process generated an initial set of preliminary codes, which were iteratively refined through constant comparison. To enhance transparency and provide an audit trail of the analytic process, Table 1 presents a summary of preliminary codes, representative excerpts, alongside their consolidation into higher-order dimensions.

<< Table 1: Summary of Preliminary Codes, Higher Order Dimensions and Representative Student Quotes>>

In the second cycle, the researchers refined and consolidated categories through constant comparison across the dataset, ensuring that themes represented distinct forms of AI–learner cognitive interaction. Disagreements in coding were resolved through discussion until consensus was achieved. To enhance the trustworthiness of the coding process, inter-rater reliability was calculated. A subset of 25% of the dataset was independently coded by both researchers, and agreement was assessed using Cohen's $\kappa$. The initial $\kappa$ was 0.78, indicating substantial agreement. Discrepancies were discussed and resolved through iterative dialogue, leading to refinement of the coding scheme. The revised coding framework was then applied to the full dataset.

Given the interpretive nature of qualitative content analysis, reliability was treated not as a fixed metric but as part of an ongoing process of negotiated meaning-making (Braun & Clarke, 2021), where consensus-building contributed to analytic depth rather than merely agreement.

The final outcome was a set of nine interrelated dimensions representing ways students described AI as participating in their thinking processes. These dimensions collectively form the AI as a Dynamic Cognitive Partner typology, grounded in learner perspectives while interpreted through theories of mediated learning and self-regulated learning.

**Findings: AI as a Dynamic Cognitive Partner Across Learning Processes**

Analysis of students' written responses reveals that learners conceptualise AI not as a single instructional tool, but as a dynamic cognitive partner whose role shifts across learning situations. Across nine interrelated dimensions, AI was described as contributing to explanation, feedback, idea generation, organisation, personalisation, monitoring, workload regulation, continuity of learning, and representational reframing. A consistent pattern across all dimensions was that students distinguished between two modes of use: AI as cognitive extension, where it supports thinking, and AI as cognitive substitution, where it replaces thinking. This boundary was not abstract but grounded in learners' own experiences of when AI helped them understand and when it reduced their engagement. Table 1 summarises these dimensions and their associated cognitive functions.

**Dimension 1: Conceptual Scaffolding**

Conceptual Scaffolding is when AI helps learners understand difficult ideas by breaking concepts into manageable steps, modelling reasoning, and explaining in ways that support the learner's own meaning-making. Students most frequently described AI as supporting understanding by breaking down complex ideas into manageable steps, modelling reasoning, and offering explanations that could be revisited and adapted. In this role, AI was positioned as enabling access to explanations that learners felt they could not reach independently: "*Photomath… gives a step-by-step explanation… helping you to understand complicated equations you can't solve" and "AI would not just give simple answers, but detailed explanations for each step… it allows me to learn and clear my questions*."

What stands out in these accounts is not just access to explanation, but control over explanation. Students emphasised being able to ask repeatedly, request clarification, and engage with explanations at their own pace. This suggests that AI is not simply delivering information but reshaping how explanation is accessed within the learning process.

At the same time, students were highly aware that this support could shift into something qualitatively different. The same features that made AI useful, immediacy, completeness, and ease, also made it easy to bypass thinking: "*AI can be detrimental… if they become overly dependent… leading to a lack of critical thinking" and "Instead of using their brains… they'll just paste the answer from ChatGPT*."

Here, the role of AI moves from supporting understanding to replacing it. Rather than working through the steps, learners can skip directly to solutions. Students described this shift in terms of passivity: "*Learning becomes passive when students let AI think for them*."

This contrast reveals a key insight: the difference is not in what AI does, but in how it is used. The same step-by-step explanation can either support sense-making or enable avoidance of cognitive effort.

Across accounts, students implicitly articulated a boundary:

- **When AI explanations are used to engage with reasoning, they extend cognition**
- **When they are used to bypass reasoning, they substitute cognition**

This distinction reflects a broader pattern across the dataset, where AI's role is not fixed but depends on how learners position it within their thinking.

**Dimension 2: Feedback & Error Detection**

This dimension refers to how learners use AI to identify errors, evaluate their work, and guide revision. AI functions as an immediate feedback source that can support or replace learners' own evaluative processes.

Students consistently described AI as supporting improvement by detecting mistakes and providing immediate feedback. Rather than waiting for teacher input, learners positioned AI as an always-available feedback partner that could identify issues and suggest revisions in real time. For example, one student explained, "*Mostly, I use Grammarly and Quillbot (AI tools) to check my grammar once I finish my draft. Those grammar checker apps can make me realise the tiny mistakes that I have made while writing my essays.*" Others described actively seeking evaluative input from AI, with one noting, "*After drafting my essays, I seek feedback from ChatGPT, asking it to evaluate my work and provide suggestions for improvement*."

What stands out in these accounts is the speed and accessibility of feedback, which allowed students to revise independently and iteratively. Feedback was not only corrective but also directional, helping learners understand how to improve rather than simply identifying errors. In this sense, AI functioned as an external evaluative lens, enabling learners to notice gaps and adjust their work accordingly.

At the same time, students recognised that this support could shift into over-reliance. When AI feedback was treated as authoritative rather than something to be evaluated, learners described a loss of engagement in the revision process. One student cautioned that "*A caveat of using an AI chatbot is that the content generated may not always be accurate and not always trustworthy*." Others observed that "*Students can just type in what they want and let AI do the work for them.*" In such cases, feedback no longer supports revision but replaces the need for learners to judge their own work, contributing to reduced critical engagement and increasing passivity, where "*Students may lose the ability to think for themselves.*"

This contrast highlights a key distinction within this dimension. When learners use AI feedback to support their own evaluation and revision, it extends their cognitive engagement. When they rely on it uncritically, evaluation is effectively outsourced, and their role in monitoring diminishes. As with conceptual scaffolding, the difference lies not in what AI does, but in how learners position it within their thinking.

**Dimension 3: Idea Stimulation**

This dimension refers to how learners use AI to generate, expand, or explore ideas, particularly at the early stages of thinking. AI can function as a source of prompts or perspectives that either support or replace learners' own idea generation.

Students frequently described AI as helping them initiate and expand ideas, especially when they faced difficulty getting started. Rather than replacing thinking entirely, AI was often positioned as a way to overcome creative blocks and open up possible directions. For example, one student noted that "*AI is best for creativity blocks… It can generate thousands of ideas, and in collaboration with our amazing human brain, millions of answers can be sought by just a single prompt.*"

What is notable in these accounts is that AI was not described as producing final answers, but as broadening the space of possibilities. Students used it to generate starting points, alternative perspectives, or rough structures that they could then develop further. Several explicitly maintained ownership over the process, emphasising that AI-supported ideation still required their own contribution. As one student explained, "*AI gives ideas but I still develop them myself,*" while another noted, "*I sometimes use AI to generate layouts or mild suggestions… but I don't rely on it to complete all my work*."

In this sense, AI reduced the barrier to initiating tasks, allowing learners to move more quickly from uncertainty to exploration. By providing initial prompts, it helped shift cognitive effort away from "blank-page" difficulty toward elaboration and refinement.

At the same time, students clearly recognised that this support could shift into over-reliance. When AI was used not to stimulate thinking but to replace it, learners described a loss of creative engagement. One student cautioned that "*Students may become overly dependent on AI tools, which could hinder their ability to think critically,*" while another observed that "*Students just ask AI to create layouts instead of struggling productively.*" In these cases, the process of generating ideas, an essential part of creative thinking is bypassed.

Some students further expressed concern that frequent reliance on AI-generated ideas could shape how they approach thinking itself, warning that "*creativity could be suppressed if AI always gives ideas first*." This highlights a shift from using AI as a starting point to depending on it as the primary source of ideas.

Across these accounts, a clear distinction emerges. When learners use AI to stimulate and extend their own thinking, it functions as a support for creative exploration. When they rely on it to generate ideas in place of their own effort, ideation is effectively outsourced. As with other dimensions, the difference lies not in the capability of AI, but in how learners position it within their thinking.

**Dimension 4: Cognitive Organisation**

This dimension refers to how learners use AI to structure and manage information, such as summarising content, creating outlines, and highlighting key ideas. AI can support or replace the cognitive work involved in organising knowledge.

Students frequently described AI as helping them manage and structure complex information, particularly during revision and study. Rather than simply retrieving answers, AI was used to transform large amounts of content into more accessible forms. For example, one student explained that *"AI can distil complex chapters into key points, making it easier for me to grasp essential concepts during revision,"* while another noted that *"it can generate step-by-step guides… and even give us summarisations of notes."* Others highlighted its role in handling information overload, with one student stating that *"AI summaries help me manage information overload when my notes are too much."*

What is notable in these accounts is that AI was used to restructure information rather than generate it, allowing learners to engage with content in more manageable forms. By condensing and organising material, AI reduced the effort required to locate and prioritise key ideas, enabling students to focus more on understanding relationships and main concepts. This function was also evident in study practices, where students described using AI to create learning materials, such as when one noted, *"I use AI to create flashcards and quizzes… It's a quick and effective method of studying."*

However, students also recognised that this support could shift into over-reliance. Several expressed concern that when AI takes over the process of organising information, learners may engage less deeply with the material. As one student reflected, *"If AI summarises everything, I might not process deeply myself,"* while others observed that *"students rely on AI notes instead of making their own"* and that *"information is consumed quickly but not retained."*

These accounts point to a critical distinction. Organising information is not only a means of managing content but also a cognitive process through which learners construct understanding. When AI is used to support this process, it helps learners navigate complexity and focus on key ideas. When it replaces the process entirely, learners may bypass the effort required to structure knowledge for themselves.

Thus, within this dimension, AI can function either as a tool that supports meaningful organisation of knowledge or as a mechanism that compresses information in ways that reduce engagement, depending on how learners use it.

**Dimension 5: Adaptive Tutoring Support**

This dimension refers to how learners use AI to receive personalised explanations, tasks, and guidance that adjust to their level and pace. AI functions as a responsive support system that can either extend learning or shift instructional authority.

Students frequently described AI as adapting to their individual needs, positioning it as a responsive tutor that adjusts difficulty, pacing, and explanation style. Rather than conforming to fixed instruction, learners emphasised that AI could be tailored to their current level. For example, one student explained that *"By informing Code Helper… of my current coding knowledge, I can receive exercises that match my skill level, allowing me to progressively challenge myself."* Others highlighted personalisation more broadly, noting that *"AI can be very useful in personalising a student's learning experience… match each students' level of understanding and… provide customised learning materials,"* while another shared that *"AI platforms adapt lessons to my pace and explain again when I don't understand."*

What is notable in these accounts is that students experienced AI not simply as providing content, but as adjusting support in response to their needs. This allowed them to engage with tasks that were neither too easy nor too difficult, supporting progression without overwhelming them. In this sense, AI helped sustain engagement by calibrating challenge and providing repeated or alternative explanations when needed.

However, students also drew a clear boundary around the role of AI in relation to human teaching. While they valued personalised support, they expressed concern about AI taking over roles traditionally associated with teachers. As one student stated, *"The position of independent learning and teaching by teachers should never be replaced,"* while others emphasised that *"AI should support teachers, not replace them,"* and pointed out that *"Whereas a chatbot doesn't. It has no emotions…"*

These accounts highlight a distinct form of tension in this dimension. Unlike earlier dimensions where the risk was cognitive substitution, here the concern centres on the displacement of human guidance and relational support. Students valued AI's ability to personalise learning but did not see it as a substitute for the social and emotional dimensions of teaching.

Thus, adaptive tutoring was positioned as beneficial when it supplemented human instruction and supported individual progression, but problematic when it was perceived as replacing teacher involvement or redefining authority in learning.

**Dimension 6: Metacognitive Monitoring Support**

This dimension refers to how learners use AI to become aware of their learning progress, strengths, and weaknesses. AI can support or replace the processes through which learners monitor and evaluate their own understanding.

Students described AI as helping them reflect on their own learning, particularly by identifying gaps, tracking progress, and making performance more visible. Rather than providing answers directly, AI was seen as offering insights into how well they were learning. For example, one student explained that *"AI can analyse a student's learning patterns and preferences… track progress… identifying areas where a student excels and where they struggle."* Another noted that *"Using the power of AI, teachers can use it to monitor the learning patterns of students… It would be certainly easier to monitor 30 students with an AI companion than mere sheets of paper."*

What stands out in these accounts is that AI was positioned as a form of externalised awareness, making aspects of learning that are typically internal, such as progress and understanding—more visible. This allowed learners to identify weaknesses, adjust their strategies, and plan next steps more effectively.

At the same time, students recognised limitations in relying on AI for this function. Some highlighted the importance of understanding how and when to trust AI-generated insights, noting that *"It is essential for students to know the limits and understand the ethicality of using artificial intelligence."* Others pointed out that AI may not capture important aspects of learning that human teachers can observe, with one student stating that *"AI assumes I understand if I don't ask questions — teachers can see from my behavior."*

These concerns point to a key risk within this dimension: the possibility of illusory understanding. When learners rely on AI-generated indicators without actively reflecting on their own thinking, monitoring can become superficial. Instead of developing self-awareness, learners may depend on external signals to judge their progress.

This suggests a distinction between two forms of use. When AI is used to support reflection and guide self-evaluation, it strengthens learners' awareness of their own learning. When it is relied on as the primary source of evaluation, self-monitoring is effectively outsourced, and the development of metacognitive skills may be weakened.

**Dimension 7: Task & Cognitive Load Regulation**

This dimension refers to how learners use AI to manage tasks, organise study processes, and reduce feelings of overload. AI can support or replace the effort involved in planning, organising, and persisting through learning tasks.

Students frequently described AI as helping them manage the demands of learning, particularly by structuring tasks, organising schedules, and reducing feelings of overwhelm. Rather than directly providing answers, AI was positioned as supporting how work is carried out. For example, one student explained that *"AI programs like Trello (AI tool) can help me plan out my study time, remind me of impending deadlines, and divide complicated jobs into smaller, more manageable chunks,"* while another noted that *"AI reminders help me stay organised and manage deadlines."*

What stands out in these accounts is that AI was used to coordinate learning activity, helping students break down tasks and maintain progress. By externalising planning and organisation, learners were able to focus more on completing tasks rather than managing them. In this sense, AI reduced the burden of organising work, making it easier for students to engage with learning tasks consistently.

However, students also recognised that this support could shift into over-reliance. When AI was used not to manage tasks but to complete them, learners described a loss of engagement with the learning process. As one student noted, *"Students can also ask AI… to help them finish their homework… and they cannot learn in the process,"* while another reflected that *"Because it is convenient, it is easy to over rely on it."*

This reveals a key distinction within this dimension. Managing tasks is part of the learning process, but when convenience becomes the priority, learners may use AI to bypass effort rather than organise it. In such cases, AI no longer supports persistence but enables avoidance.

Thus, AI can function as a tool that supports task management and sustained engagement, or as a means of escaping the cognitive effort that tasks are designed to provoke, depending on how learners use it.

**Dimension 8: Learning Continuity Support**

This dimension refers to how learners use AI to continue learning beyond classroom boundaries. AI provides ongoing access to support, enabling learners to sustain learning across time and contexts.

Students frequently described AI as enabling them to continue learning outside formal classroom settings, particularly when teachers or peers were not available. AI was positioned as a constant source of support that allowed learning to extend beyond scheduled instruction. For example, one student explained that *"When I don't have anyone to teach me, AI helps. It is very efficient and convenient,"* while another noted that *"AI is available anywhere and anytime so I can still not miss materials,"* and similarly, *"AI is available when teachers are not."*

What is notable in these accounts is that AI was not only valued for what it provides, but for when and where it is available. Students used it to maintain continuity in their learning, accessing explanations and support at moments of need rather than waiting for formal instruction. This allowed them to sustain engagement, particularly when encountering difficulties outside class time.

However, students also recognised that this continuity could come at a cost. Several expressed concern that reliance on AI might reduce interaction with teachers and peers. As one student observed, *"Students may stop asking teachers… Teacher–student relationships weaken,"* while another noted that *"Less human interaction harms social development."* Others pointed to the ease of dependence, warning that *"Because it is convenient, it is easy to over rely on it."*

These accounts highlight a tension between continuity and isolation. While AI extends access to learning, it may also narrow the social dimensions of learning if it becomes the primary source of support. Instead of expanding participation in a learning community, interaction may shift toward individual engagement with a tool.

Thus, AI can function as a resource that extends learning across time and context, supporting persistence and access, but it may also contribute to reduced human interaction and relational engagement when relied on as a primary substitute for teachers or peers.

**Dimension 9: Explanation Reframing**

This dimension refers to how learners use AI to transform how information is presented, such as simplifying language, providing alternative explanations, or offering different examples. AI can support or replace the effort involved in making sense of complex ideas.

Students frequently described AI as helping them reframe explanations, allowing them to encounter the same concept in different forms, languages and modals. Rather than introducing new content, AI was used to adjust how information was presented, making it more accessible or easier to understand. For example, learners reported asking AI to simplify explanations, such as requesting that content be *"explained as if I were a 10-year-old,"* or using it to convert dense material into language that is *"easier to read."* Others noted that AI *"explains in different ways until I understand,"* provides alternative examples, and helps overcome language barriers through translation.

What stands out in these accounts is that AI supports understanding by changing the representation of knowledge, enabling learners to approach concepts from multiple entry points. By rephrasing and restructuring explanations, AI helps learners overcome barriers in comprehension and continue progressing when initial explanations are unclear.

At the same time, students were aware that this flexibility could introduce risks. Several noted that AI-generated explanations are *"not always accurate,"* may contain *"errors,"* or produce *"weird steps and wrong answers in math,"* requiring verification. More importantly, students expressed concern that simplified explanations might lead to shallow understanding. When information is made easier too quickly, learners may accept it without fully processing the underlying concepts.

This reveals a key distinction within this dimension. Reframing can support comprehension when learners actively engage with alternative explanations and use them to build understanding. However, when simplification replaces effort, learners may develop an illusion of understanding, where clarity is achieved without depth.

Thus, AI can function as a tool that supports meaning-making through multiple representations, or as a mechanism that reduces conceptual engagement through over-simplification, depending on how learners use it.

It is important to note that the nine dimensions are analytically distinct but often overlap in practice. Students' accounts frequently reflected multiple cognitive functions at the same time, for example, feedback from AI could also support monitoring, or scaffolding could involve re-explaining ideas in different ways. This reflects the reality that learning processes are interconnected rather than neatly separated.

For this reason, the dimensions should not be understood as fixed or mutually exclusive categories. Instead, they represent the main role AI plays in a particular moment of learning. During analysis, each segment was coded based on its primary function, while recognising that other functions could also be present. This approach allowed us to organise the data clearly without losing the complexity of how AI is actually used.

To keep the dimensions conceptually clear, we paid particular attention to distinguishing closely related functions. For example, *feedback and error detection* (Dimension 2) refers to AI identifying mistakes or providing suggestions, whereas *metacognitive monitoring support* (Dimension 6) refers to how learners interpret that feedback to understand their own progress. Similarly, *cognitive organisation* (Dimension 4) focuses on structuring information (e.g., summaries or outlines), while *task and cognitive load regulation* (Dimension 7) concerns managing workload and effort. In addition, *conceptual scaffolding* (Dimension 1) involves step-by-step support for understanding, whereas *explanation reframing* (Dimension 9) involves presenting the same idea in different ways.

Overall, while the dimensions are analytically separated for clarity, students' experiences and perception show that AI-supported learning is fluid, overlapping, and shaped by context.

**Discussion**

The typology's contribution is not simply a list that "lists" nine learner-described uses of AI, but that it reconceptualises what counts as AI's educational role and where learning is located when learners study with GenAI systems. First, it shifts the unit of analysis from tool categories (e.g., tutor, chatbot, grader) toward learner-attributed cognitive functions, aligning with learning-sciences traditions that treat technologies as mediational means whose meaning emerges in use rather than in design labels (Vygotsky, 1978). Second, it theorises AI use as dynamic role shifting across phases of activity, consistent with activity theory's claim that tools reconfigure action in relation to goals, rules, and division of labour (Engeström, 1987), and with SRL's phase models in which supports are recruited differently during planning, monitoring, and reflection (Zimmerman, 2000). Third, and most importantly, it foregrounds a support–substitution boundary as a core theoretical problem: the same AI interaction can either extend cognition (scaffolding, reorganising representations, reducing extraneous load) or replace cognition (outsourcing sense-making, metacognitive laziness and AI dependent thinker). This boundary reframes "overreliance" from a purely ethical or policy issue into a psychological and interactional phenomenon: a shift in epistemic agency and responsibility for thinking within mediated activity. In doing so, the typology integrates sociocultural mediation, distributed cognition, SRL, and cognitive load theory into a coherent account of how AI becomes part of the learners' cognitive systems while also creating conditions for cognitive offloading that may undermine learning.

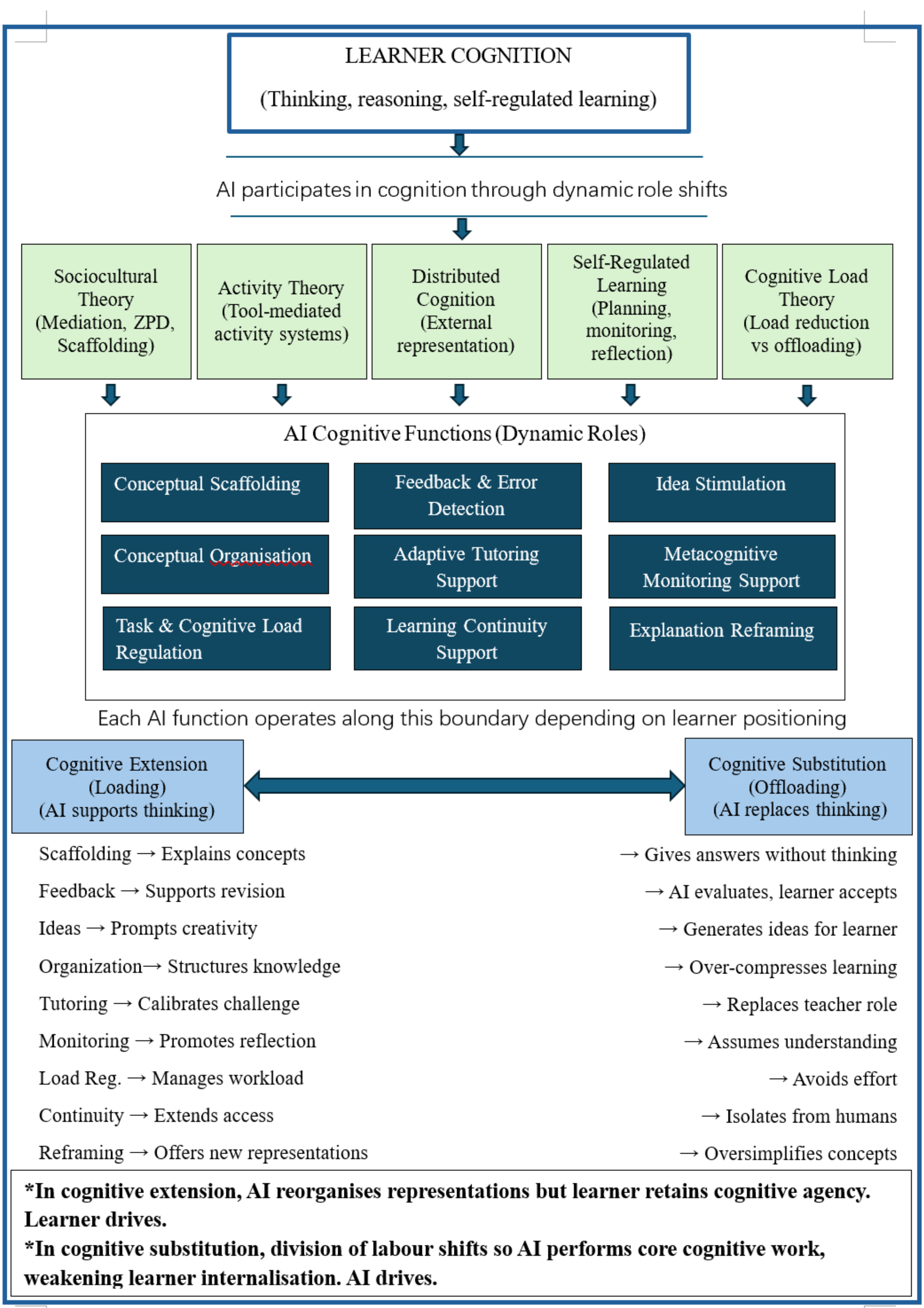

Figure 1: AI as Dynamic Cognitive Partner Framework

**A Typology for AI as a Dynamic Cognitive Partner**

Figure 1 operationalises the Dynamic Cognitive Partner claim by depicting AI as an agent that enters learners' thinking through dynamic role shifts rather than a single, fixed "tool role". At the top, Learner Cognition anchors the typology in students' reasoning and self-regulated learning, while the central box specifies the nine AI cognitive functions that learners reported recruiting across study situations (e.g., conceptual scaffolding, feedback/error detection, idea stimulation, organisation, adaptive tutoring, metacognitive monitoring, task/load regulation, learning continuity, and explanation reframing). Critically, the typology adds an evaluative boundary condition: each function can be enacted along a continuum from cognitive extension ("AI supports thinking") to cognitive substitution ("AI replaces thinking"), depending on learner positioning, that is, whether the learner retains epistemic agency for sense-making, evaluation, and decision-making, or delegates these core processes to AI. This boundary reframes "overreliance" as a situated shift in the division of cognitive labour, not merely the amount of AI use. The theoretical foundations at the base clarify why the same interaction can be productive or harmful: sociocultural mediation and scaffolding explain contingent support; activity theory explains shifting roles within goal-directed activity systems; distributed cognition explains external representation and coordination; SRL explains phase-specific regulation demands; and cognitive load theory explains when load reduction becomes offloading.

**1. From "what the tool is" to "what the learner does with it" (a mediational turn)**

The first theoretical move is a mediational redefinition of AI's role. Sociocultural theory argues that higher mental functions develop through mediated action with cultural tools and signs (Vygotsky, 1978). Crucially, the "role" of a tool is not fixed by its material form; it is enacted in activity. This aligns with activity theory that agency is distributed across people and mediational means, and that the same artifact can enable different forms of action depending on how it is taken up. By organising AI use into learner-attributed cognitive roles (scaffolding, monitoring, organisation, persistence support, etc.), the typology operationalises this insight for the use of GenAI in the course: AI is theorised not as a "tutor" in the abstract but as a mediational means recruited for specific cognitive functions. This matters because "AI tutor" or "chatbot" categories can obscure important distinctions: a chatbot can function as a conceptual scaffold in one moment and as a shortcut in another. The typology therefore pushes theory and research away from platform labels toward the psychological function of interaction, which is closer to how mediation and learning are theorised in the learning sciences.

Prior research (Gerlich, 2025; Watson & Ennion, 2026) has shown that effects of AI vary depending on whether AI was used for *sense-making* (supportive mediation) versus *answer acquisition* (substitutive offloading). This typology provides a theoretical vocabulary for explaining heterogeneity that outcome-only studies often report but cannot interpret.

**2. Dynamic role shifting as an activity-theoretic and SRL phenomenon**

SRL models conceptualise learning as cyclical phases, forethought/planning, performance/monitoring, reflection/adaptation (Zimmerman, 2000). Each phase has different regulatory demands, and learners strategically recruit supports differently across phases. The typology's dimensions map naturally onto these phase demands (e.g., idea stimulation and task chunking as planning supports; feedback and monitoring as performance supports; reflection through identifying weak points). The deeper insight is that GenAI is not only a "resource"; it can become a co-regulator, shaping learners' strategy selection, effort regulation, and monitoring frequency.

At the same time, the typology highlights a theoretical risk: when AI supplies too much regulation (e.g., it identifies weaknesses, decides next steps, evaluates correctness), the learner may shift from self-regulation toward delegated regulation, weakening metacognitive skill development. This is a sharper, more theoretically grounded reframing than the broad claim that "students rely too much on AI." It identifies which part of regulation is displaced (monitoring, evaluation, planning) and thereby provides leverage for both research and design. In fact, some students turn to AI precisely because they lack confidence in their own ability to complete the task, believing that the AI will produce superior work, their AI guilt emerges (Chan, 2025c).

**3. Clarifying the boundary between cognitive extension and cognitive substitution**

From distributed cognition, the core claim is that cognitive work can be distributed across artifacts, representations, and social partners (Hutchins, 1995). From the extended mind thesis, external resources can become functionally integrated into thinking when they are reliably available and tightly coupled to the user's goals (Clark & Chalmers, 1998). GenAI is unusually "extendable" because it is conversational, always available, and can generate bespoke representations on demand. The typology advances these perspectives by specifying how learners experience the extension: not only memory support (like notes) but explanation generation, monitoring cues, and motivational bridges.

However, distributed/extended cognition has always carried an implicit question: when does distributing cognition stop being productive and start undermining learning? Education depends on internal changes, schema construction, conceptual reorganisation, strategy acquisition, not only task completion. The typology makes this boundary visible empirically and theoretically. It shows that learners themselves articulate the difference between using AI to support sense-making versus using AI to avoid sense-making. That boundary is the point where mediation fails to become internalised in Vygotskian terms, and where the division of labour shifts so far toward the tool that the learner's epistemic agency is reduced.

Cognitive load theory helps explain why this boundary is so slippery. CLT distinguishes extraneous load (wasteful processing) from germane load (processing that contributes to learning) (Sweller, 1994). AI can reduce extraneous load by simplifying language, organising information, and providing worked examples. But it can also reduce germane load if learners skip explanation generation, error analysis, or retrieval effort, processes that are difficult precisely because they build durable knowledge structures. The typology therefore provides a theoretical mechanism for what the students described as "information consumed quickly but not retained": AI may optimise fluency and speed while inadvertently reducing the cognitive work required for schema formation. This is not a claim that AI is inherently harmful; it is a theory-driven account of a specific failure mode: cognitive bypass.

**4. Introducing "learner positioning toward AI" as an epistemic-agency construct**

The findings reveal an implicit construct: learner positioning toward AI. In the literature, "trust calibration," "AI literacy," and "help-seeking" are often treated separately. Trust calibration means knowing exactly when to trust AI and when not to trust it, so students rely on it at the right times. AI literacy means having the basic knowledge and skills to understand how AI works and how to use it properly. Help-seeking means the habit of turning to AI for assistance when doing schoolwork. The typology unifies them through a positioning lens: learners decide whether AI is an assistant to their cognition or a replacement for their cognition. This is fundamentally a question of epistemic agency, who is responsible for generating, checking, and owning the knowledge. The typology thus offers a way to theorise overreliance not simply as "too much usage" but as a shift in epistemic stance: from learner-as-author to learner-as-consumer.

This matters because it suggests new research predictions: substitutive positioning should correlate with weaker transfer, poorer metacognitive calibration, and lower resilience when AI is unavailable (e.g., exams), whereas supportive positioning should correlate with improved strategy repertoire and stronger conceptual understanding. It also provides a design lever: systems can be built to nudge learners toward supportive positioning (prompting justification, comparison, reflection) rather than substitutive positioning (instant final answers).

**Implications of the Typology**

A theoretical contribution of the Dynamic Cognitive Partner typology is that it provides a bridging model between established learning theories and the distinctive interactional properties of GenAI. Much prior AI-in-education work either (a) evaluates outcomes such as achievement, satisfaction, time saved, acceptance (e.g., Bai & Wang, 2025; Wu & Yu, 2024; Wu, 2026), (b) categorises systems by instructional function, for example, tutor, recommender, automated scoring (e.g., Martin et al., 2024; Xu & Ouyang, 2022), or (c) focuses on ethics and governance (e.g., Borenstein & Howard, 2021; Li & Zhang, 2025). Those approaches are necessary, but they often under-theorise a key shift introduced by GenAI: learners are no longer only receiving instruction from a system; they are thinking with it in real time through dialogue, iterative prompting, and rapid re-representation. This typology contributes by explaining this shift at the level where learning actually happens: moment-to-moment cognitive activity under task demands and constraint.

A second contribution is that the typology treats differences in how students use AI as something theoretically meaningful, rather than mere noise or random variation. According to activity theory, tools mediate the relationship between the user (subject) and the task (object) within a broader activity system that is shaped by rules, community, and division of labour (Engeström, 1987).

Generative AI intensifies this because it can rapidly take on multiple functions within the same activity (e.g., a learner writing an essay can use AI to generate an outline, then ask for counterarguments, then check grammar, then request citations). The typology as "dynamic partner" claim is essentially an activity-theoretic claim: AI reconfigures the learning activity by enabling micro-transitions between sub-goals, thereby changing how the learning process is organised over time.

Finally, the typology's theoretical value lies in integration. Sociocultural theory explains mediation and internalisation; activity theory explains goal-directed systems and division of labour; distributed cognition explains external representation and coordination; SRL explains regulation cycles; CLT explains capacity constraints and offloading risks. Many papers reference one or two of these lenses, but few provide a coherent model that can simultaneously explain why AI feels empowering, why it changes learning practices, and why it can undermine learning when mispositioned. The typology functions as that integrative model, anchored in learner accounts rather than imposed categories, making it especially valuable for a field that is currently fragmented across tool types, outcomes, and policy debates.

Practically, the typology can be used (a) by educators to design AI-supported tasks that preserve learner agency by pairing each function with prompts for justification, comparison, and reflection; (b) by learners as a self-check to identify when AI is extending versus substituting their thinking; and (c) by researchers as a design heuristic, a basis for hypothesis testing, and an analytic coding scheme for characterising AI–learner partnerships across contexts.

**Conclusion**

This study reconceptualises AI's educational role through a learner-informed lens, positioning AI not as a static instructional tool but as a dynamic cognitive partner whose function shifts across learning situations. Analysis of students' written accounts shows that learners perceive AI as participating directly in thinking processes, explaining, organising, stimulating ideas, monitoring understanding, regulating workload, and sustaining learning beyond classroom contexts. Crucially, students themselves articulated a boundary between cognitive extension and cognitive substitution, demonstrating awareness that AI can either support understanding and self-regulation or replace effortful cognitive work.

The typology contributes theoretically by relocating AI from the periphery of instruction to the core of cognitive activity. It integrates sociocultural mediation, distributed cognition, activity theory, self-regulated learning, and cognitive load perspectives into a unified account of how AI becomes part of learners' cognitive systems. At the same time, it clarifies that AI's educational value is not inherent in the technology but emerges from how learners position AI within their activity as a scaffold for sense-making or as a shortcut that bypasses it. By foregrounding this support–substitution boundary, the typology reframes overreliance as a cognitive and interactional phenomenon rather than solely an ethical issue.

Several limitations should be considered. The study relies on self-reported written responses rather than measuring learning outcomes, which capture learners' perceptions and experiences of AI use rather than direct observation of behavior. Students may describe idealised or socially desirable practices, and their accounts may not fully represent the exact moment of cognitive processes during actual AI interaction. Furthermore, students' responses might have been influenced by the AI literacy course they had undertaken, meaning that their perceptions might have reflected the course curriculum and content rather than their authentic insights. The limited scope (based on an AI literacy course in the Hong Kong context), data (student perceptions in written responses), and sample of this study also restrict the generalisability of the findings.

Finally, AI systems evolve rapidly. The typology captures patterns of mediation characteristic of current conversational and generative systems, but specific interaction forms may change as technologies develop.

Future research should move from conceptual description toward design and intervention studies that support learners in maintaining epistemic agency when working with AI. One promising direction is the development of structured classroom implementations that help learners become more aware of when they are engaging in productive cognitive effort and when they are offloading core thinking processes. Such supports could include reflective prompts, metacognitive checklists, or activity routines that encourage learners to pause, evaluate their understanding, and justify AI-supported decisions. These forms of implementation would not restrict AI use, but instead cultivate learners' ability to regulate how AI is integrated into their cognitive work.

In addition, future studies should investigate how different learner positionings toward AI relate to learning outcomes, transfer, and long-term strategy development. Comparative research across age groups, subject domains, and cultural contexts would further refine the typology. Finally, combining self-report data with learning analytics, screen capture, or think-aloud protocols could provide richer insight into the real-time dynamics of human–AI cognitive partnerships.